\documentclass[preprint,prb,showpacs,amsmath,amssymb,floatfix]{revtex4}
\usepackage{graphicx}

\newcommand{\bn}{\begin{equation}}
\newcommand{\ee}{\end{equation}}
\newcommand{\bga}{\begin{eqnarray}}
\newcommand{\eda}{\end{eqnarray}}

\newcommand{\diff}{\mathrm{d}}
\newcommand{\eps}{\epsilon}

\newcommand{\bigO}{\mathcal{O}}

\newcommand{\chalmersMC}{Department of Microtechnology and Nanoscience, MC2,
Chalmers University of Technology,
SE-41296 G\"{o}teborg, Sweden}

\begin{document}
\title{A general solution to the Schr\"odinger-Poission equation for charged hard wall: Application to potential profile of an AlN/GaN barrier structure}

\author{Kristian Berland} \affiliation{\chalmersMC}
%\pacs{73.20.-r}{Electron states at surfaces and interfaces}
%\pacs{71.20.Nr}{Semiconductor compounds}
%\pacs{74.78.Fk}{Multilayers, superlattices, heterostructures }
%\pacs{03.65.-w}{Quantum mechanics}

%\email[{berland@chalmers.se}]

\begin{abstract}
A general, system-independent formulation of the parabolic Schr\"odinger-Poisson equation is presented for a charged hard wall in the limit of complete screening by the ground state. It is solved numerically using iteration and asymptotic-boundary conditions. 
The solution gives a simple relation between the band bending and charge density at an interface. I further develop approximative analytical forms for the potential and wave function, based on properties of the exact solution.
Specific tests of the validity of the assumptions leading to the general solution are made.
The assumption of complete screening by the ground state is found be a limitation; however, the general solution still provides a fair approximate account of the potential when the bulk is doped.
The general solution is further used in a simple model for the potential profile of an AlN/GaN barrier, and gives an approximation which compares well with the solution of the full Schr\"odinger-Poisson equation.
\end{abstract}

\maketitle

\date{\today}

Quasi two-dimensional electron gases (2DEGs) form at many planar interfaces and surfaces where electron accumulate in inversion layers. \cite{Schrieffer,2d:ando}  
They play a central role for the operation of many devices, for instance for metal-oxide semiconductor (MOS) devices, and high-electron mobility transistors (HEMT). 
Naturally, their properties such as quantized levels and conduction band bending have been much studied both experimentally and theoretically.\cite{Fang:1967,FrankHoward,Duke,Pals:general,Gwo:ClassicalAndQuantum,Appelbaum:Efield,King:InO3,Hyldgaard:Sic}  
In particular the angle-resolved photoemission spectroscopy (ARPES) characterisation of the InN surfaces has spurred recent activity. \cite{InN:main,King:narrow,King:SP,King:universal}
Heterojunctions of highly polar materials, such as the III-V nitrides,\cite{2DEG:AlN} induces these 2DEGs at the positively charged interfaces. 
A good account of the band bending at interfaces in these materials is essential for band-gap engineered intersubband devices such as resonant-tunneling diodes and quantum-cascade lasers. Simple quantum-mechanical systems, such as the particle in box, harmonic oscillator, and linear potential well are instructive model useful to generate rough accounts of various physical phenomena described by the Schr\"odinger equation. 
In the same vein, the charged hard wall represent a model case for the Schr\"odinger-Poission (SP) equation describing the quantization and band bending at interfaces.

The conduction-band edge, or potential, $V$ and quantized levels $E_n$ 
at interfaces are usually obtained with the SP equation with mass $m$, dielectric constant $\epsilon$, 
\begin{equation}
\left[-\frac{\hbar^2}{2 m} \frac{\diff^2}{ \diff z^2 }  + V(z) \right] \psi(z) = E \psi(z)\,, \quad 
-\epsilon\frac{\diff^2}{\diff z^2} V(z) = \rho (z)  \\
  \label{eq:SP}\,,
\end{equation}
where $\rho(z)$ is the total charge-density comprised of donor, interface, and electron charge.
The related textbook linear-potential well problem is inappropriate because 
it lacks an account of the electron screening inherit to the problem.
The SP equation is usually solved iteratively; $V$ is updated until it reaches self-consistency. 
%&	&\psi(0)=\psi(z\rightarrow \infty)=0, \quad \,. \label{eq:Gauss}
This approach is straightforward to implement, but as a first line of attack to device modelling and for understanding physical trends, simple analytical results are also of great value.

In this paper, a general, system-independent, formulation of the parabolic Schr\"odinger-Poisson equation for a charged hard wall is presented in the limit of complete screening of the interface charge by the ground-state, that is, in the quantum electrical limit.\cite{Appelbaum:Efield,Pals:general} It is solved numerically using iteration. 
These steps follow the earlier work of Pals,\cite{Pals:general} who 
also provided an analytical approximation using the variational principle. 
In contrast to Pals, I present analytical expressions that are based on constraints from physical principles and exact properties obtained from the numerical solution.
Furthermore, I make specific tests of the general solution outside its expected range of validity. 
Finally, I demonstrate the usefulness of the analytical expressions for 
making simple models of the band bending in AlN/GaN heterostrucutes. 

A key assumption made to arrive at the model system is the infinite potential barrier or hard wall at $z=0$. It is appropriate for flat surfaces, as the potential variation is abrupt and the work function is much larger than other characteristic energies; for interfaces, the band offset must be large. Another, is the neglect of non-parabolicity, which is an important effect in some semiconductors, but more-so for excited states of narrow quantum wells than for the ground-state of the shallow quantum wells that form at interfaces.

The assumption of complete screening of the interface charge $\sigma$ by the ground state $\psi_0$ leads to 
\begin{equation}
	\rho(z)=\sigma[\delta(z)- |\psi_0(z)|^2]\,,
\end{equation}
with $\sigma=m(\epsilon_F-E_0)/\pi$ for an isotropic 2DEG in zero magnetic field. This assumption is a serious limitation, as it is both a zero-temperature ($T=0$) condition and a restriction on the amount of charge at the interface. 
For $T=0$, it is valid when the Fermi level is below the first excited state.
\begin{figure}[t]
\includegraphics[width=8cm]{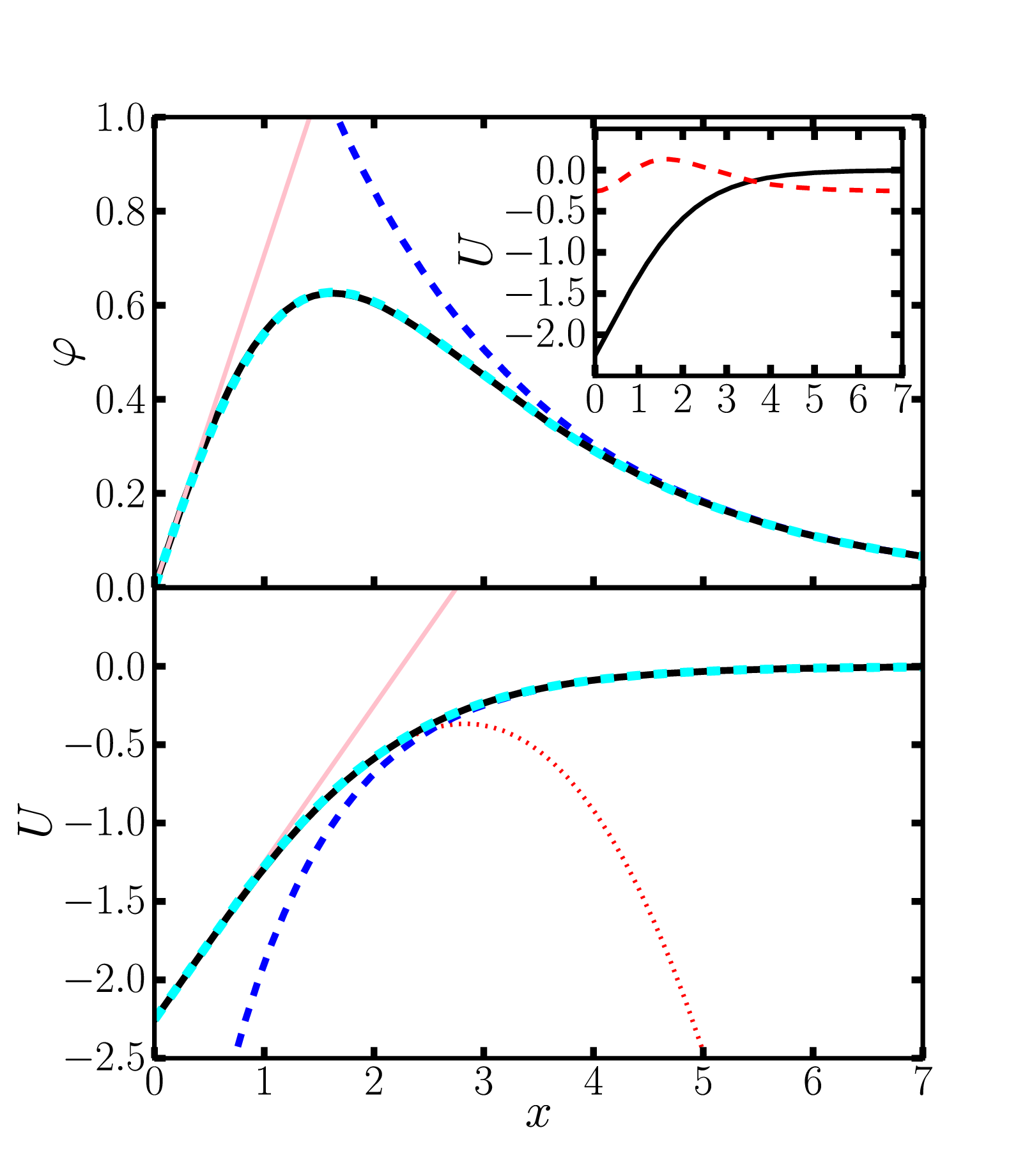}
\caption{Numerically determined wave function and potential for the system-independent formulation of the Schr\"odinger-Poisson equation for a charged hard wall. 
The full dark line in the  upper (lower) panel gives the wave function (potential). 
The dashed dark [blue] lines show the asymptotic curves, while the thin light [pink] lines show the first (second) order expansion of wave function (potential). 
The dotted line gives the third order expansion of the potential . 
The dashed light [cyan] curve gives analytical approximate forms based on constraints. 
In the insert, the [red] dashed curve gives the potential and the density of the wave-function as offseted by the eigenvalue $K$.}
  \label{fig:Fig1}
  \end{figure}

The dimensionless equation that leads to the general solution are obtained with the change of variables:
$E = \gamma K$, $z=\lambda x$, and $\phi=\psi \sqrt{\lambda}$. Here  $\lambda = \left(\hbar^2 \epsilon/2m \sigma \right)^{1/3}$ and $\gamma=\lambda \sigma /\eps$ defines length and an energy scales. 
The two first change of variables are identical to the textbook procedure for a linear potential well. \cite{Mahan:Nutshell}
We get
\begin{eqnarray}
	&&\left[ - \frac{\diff^2}{\diff x^2}  + U(x) \right]   \phi (x) = K \phi(x) \nonumber
  \\ 
	&&U(x) =  \left[ x - \int_0^x \diff x'\, (x-x')|\psi(x')|^2 -<x>\right]\,,  \label{eq:main}
 \end{eqnarray}
and note that we need only solve this equation once. The system-specific wave function $\psi$, potential $V$, and ground state eigenvalue $E_0$ can be restored for specific values of $\sigma$, $m$, and $\epsilon$.

To guide the computational procedure, I first consider certain limits. For large $x$, $|K|>>|U|$, and the asymptote of the wave function follows $\phi \rightarrow  A_\phi \exp[-\sqrt{-K} x]$, which further leads to the asymptote of the potential $U\rightarrow A_U  \exp[-g x]$, where $A_U=A_\phi^2/4K$ and $g=-2\sqrt{-K}$ is the decay factor. 
For small $x$, the wave function is linear to first order $\phi(x)=\zeta x +\bigO(x^2)$, with $\zeta=\phi'(0)$, and the first three terms in the expansion of the potential ensues
\begin{equation}
U(x)=U_0 + x - \frac{\zeta^2}{12} x^3+\bigO(x^4)
    \label{eq:pot-terms}\,.
\end{equation}
We can also identify $U_0=-<x>$. 

The numerical solution of Eq.~(\ref{eq:main}) is obtained using iteration, similar to the solution of the full SP equation. 
For a given potential $U(x)$, the Schr\"odinger part is discretized, with uniform grid-spacing $\Delta x$ according to the finite difference method, and the eigenvalues and eigenvectors are determined with a banded eigenvalue solver.\cite{scipy} 
$K$ equals the minimal eigenvalue and the its normalized eigenvector gives the wave function $\varphi$.
Next, using this result as input, the potential $U$ is updated until self-consistency is reached. Potential mixing secures convergence.
%The loop is endes once $K$ change by less than $10^{-9}$.
To improve accuracy and simplify extraction of parameters, I use {\it asymptotic boundary conditions} (abc): 
Since, the wave function falls of exponentially, a hard wall boundary condition at $x=L$ for some large cutoff length $L$ is commonly used; however, since only a single wave function is retained, the condition $\phi(L+\Delta x)=\exp(-\sqrt{K} \Delta x) \phi(L)$ can be adopted with $\Delta x$ being the grid spacing.
$K$ is updated alongside the potential $U$ in the iterative loop. 
Unlike the hard wall condition, abc guarantees asymptotic behavior at the boundary and $\phi(L)$ can be used to obtain $A_\phi$.

\begin{table}[h]
\begin{ruledtabular}
\caption{Convergence study of numerical solution. $L$ is the length of the unit cell, $\Delta x$ is the grid spacing. 
The table give $(1-K(L,\Delta x)/K_{\rm conv})\cdot10^{-6}$ where I set $K_{\rm conv}=K(50,1/400)$.}
\begin{tabular}{lllllll} 
\hline$ L\backslash h$   &1/10 & 1/50&1/100 & 1/200&1/300 & 1/400  \\
30 &2973&442.1&196.2&79.25&41.12&22.22 \\ 
40 &2414&330.8&140.6&51.46&22.60&8.329 \\
50 &2080&264.1&107.3&34.79&11.48& 0 \\
\hline
\end{tabular}
\label{tab:conv}
\end{ruledtabular}
\end{table}

Table~\ref{tab:conv} shows the result of the convergence study. A grid spacing of $\Delta x=1/400$ and a length of $L=50$ converges $K$ within $10^{-5}$.
\begin{table}[ht]
\begin{ruledtabular}
\caption{Parameters extracted from the numerical solution (described in text), and corresponding relations to results for specific parameters. $c$ and $d$ are parameters of the constrained-based approximate fit.}
\begin{tabular}{lll}
Relation &Parameter &  Value  \\
$\psi'(0)=\sqrt{\lambda} \zeta $ &$\zeta$& 0.70708 \\
$V'(0)=U'(0)\gamma/\lambda $ &$U'(0)$& 1 \\
$z=\lambda x $&$<x>$&  2.2543  \\
$E=\gamma K$ &$K$ &  -0.25902  \\
$g=2\sqrt{-K}$ &$g$ & 1.0179  \\
$V(0)=\gamma U_0$ &$U_0$&  $-<x>$ \\
$A_{V} = \gamma A_U$ & $A_U$ & -5.2444\\
$A_{\psi}=\sqrt{\lambda} $ & $A_\phi$ & 2.3310\\
\hline
$c=\left(1+\frac{1}{U_0 g}\right)\frac{A_U}{A_U-U_0} $ & c & 0.98957 \\
$- $ & d & 1.4256  \\
\end{tabular}
\label{tab:main}
\end{ruledtabular}
\end{table}

Figure~\ref{fig:Fig1} displays the general, system-independent wave function $\phi$ and potential $U$, while table~\ref{tab:main} summarizes key parameters.
There is only a single bound state; its eigenvalue is $K$.
The value of $U_0$ gives a general relation between between the charge at the interface $\sigma$ and band banding at $x=0$:
$V(0) = U_0 \left(\hbar^2 \epsilon/2m \sigma \right)^{1/3} \sigma^{2/3}$.

\begin{figure}[t]
\includegraphics[width=8cm]{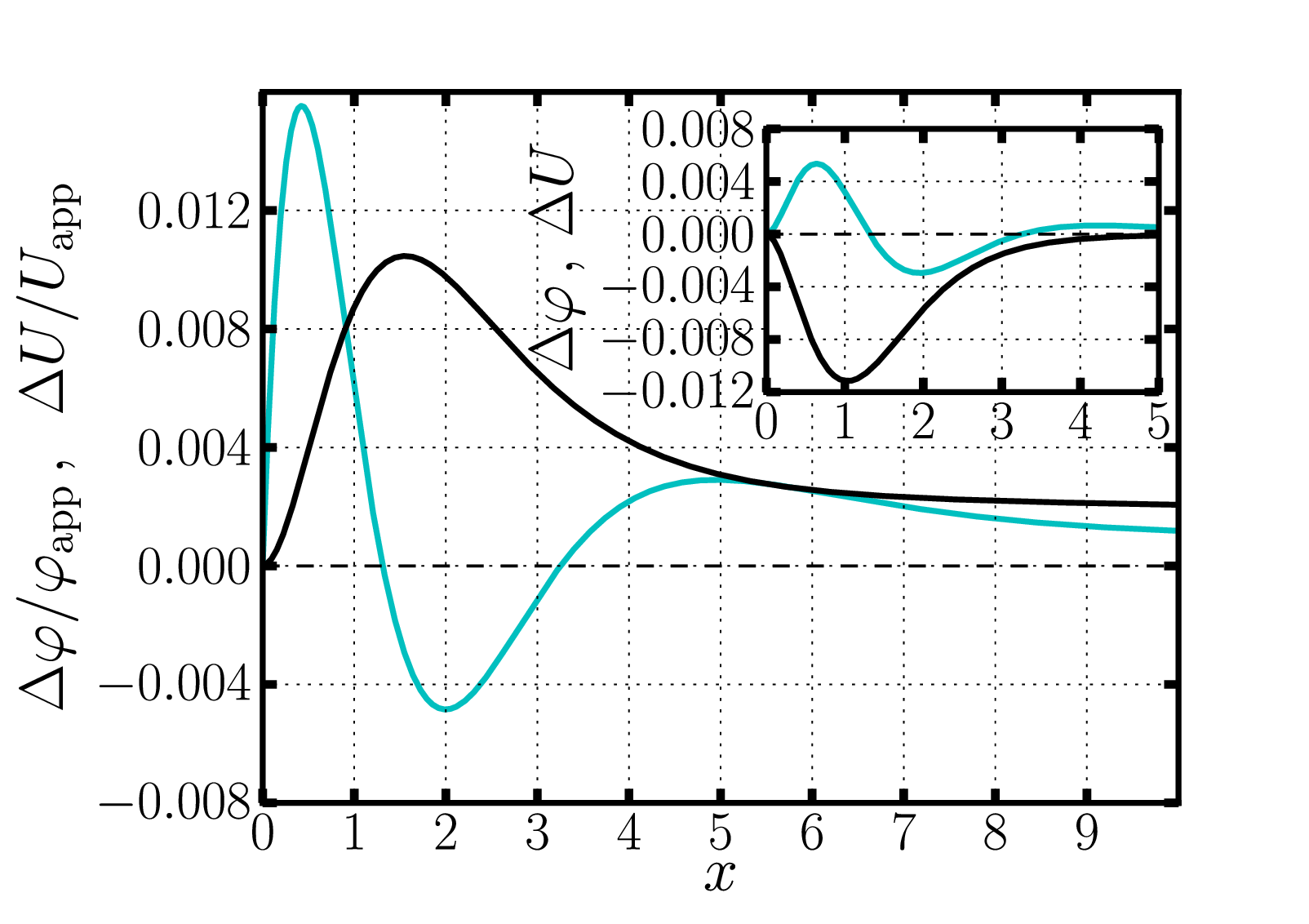}
\caption{Relative difference between the constraint-based approximation and numerical solution for potential $U$ and wave function $\varphi$. 
The dark [black] curve gives $(U-U_{\rm app})/U_{\rm app}=\Delta U / U_{\rm app}$,
while the light [cyan] curve gives $(\varphi-\varphi_{\rm app})/\varphi_{\rm app}=\Delta \varphi /\varphi_{\rm app}$. 
The insert shows absolute differences.}
\label{fig:diff}
\end{figure}

Modelling of semiconductor surfaces and interfaces can benefit from analytical approximative expressions of the wave function $\varphi_{\rm app}$ and potential $U_{\rm app}$. 
I here present such expressions, which are based on constraints stemming from the numerical solution and physical principles.
For small $x$, I choose the conditions $U_{\rm app}(x) =U_0+x +\bigO(x^2)$, and $\phi_{\rm app}(x)=\zeta x+\bigO(x^2)$, where $\zeta=\phi'(0)$,
while for large $x$ the approximative expressions should have the exact asymptote.
Moreover the wave function $\phi_{\rm app}(x)$ should be normalized, which follows from charge neutrality. 
The nodeless shape of the wave function and potential in Fig.~\ref{fig:Fig1} motivates simple expressions: 
\begin{align}
U_{\rm app}(x) &=\frac{A_U U_0 e^{-g x}}{U_0+(A_U-U_0)e^{-gcx}}\,, \\
\varphi_{\rm app}(x)&=\frac{\zeta A_{\varphi} x e^{-gx/2 } }{ \zeta x+A_{\varphi}e^{-g d x /2}}\,,
\label{eq:app}
 \end{align}
which obeys the specified conditions if $c=\left(1+1/U_0 g\right)\left[A_U/(A_U-U_0)\right]$  and $d$ is adjusted to normalize the wave function. 
Table~\ref{tab:main} lists the determined values of $c$ and $d$.
In Fig.~\ref{fig:Fig1} the dashed light curves give the approximative solution, which differ from the numerical by less than the width of the curves. 
Fig.~\ref{fig:diff} details the relative difference: $(U-U_{\rm app})/U_{\rm app}=\Delta U / U_{\rm app}$ and $(\varphi-\varphi_{\rm app})/\varphi_{\rm app}=\Delta \varphi /\varphi_{\rm app}$,
which is less than 1.6 \% for $\phi$ and 1 \% for $U$. 
The insert shows absolute differences. 
The tiny discrepancy between the approximative and the numerical solution makes the analytical expressions sufficient for most modelling purposes.
It also shows that constrained-based strategies can lead to excellent approximative expressions. 

\begin{figure}[t]
\includegraphics[width=8cm]{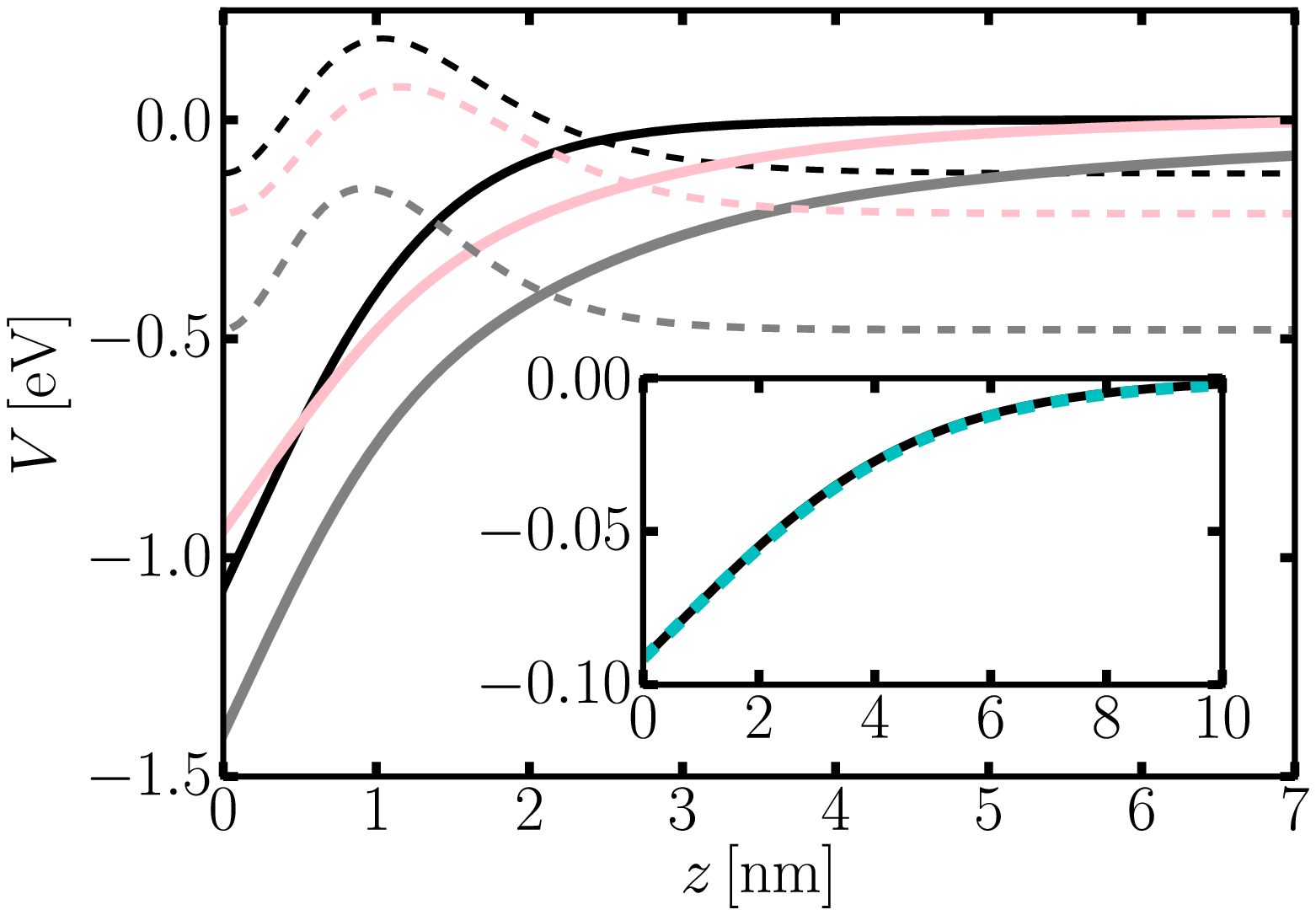}
\caption{Assessment of the general solution beyond its range of validity. 
The full curves give the potential, while the dashed indicate the wave-functions squared (in arbitrary units offseted by their energy.)
The upper [black] curves give the results obtained from the approximative form of the general solution using a sheet density of $\sigma=10^{13} {\rm cm}^{-2}$, and $m=0.2m_{\rm e}$.
The lower [gray] curves, the result obtained using a full Schr\"odinger-Poisson (SP) solver as described in Ref.~\onlinecite{berland:thermal} for the same parameters. The (middle [pink]), includes bulk doping of $\rho_{\rm d}=10^{19} {\rm cm}^{-3}$. 
The insert compares the full SP result with that based on the general solution for $\sigma=10^{12} {\rm cm}^{-2}$ at zero temperature. 
}
  \label{fig:Fig3}
 \end{figure}
The general solution has a limited range of validity. However, it can still serve as an approximate account of band bending at surfaces and interfaces capturing essential trends and as a building block in simple models. 
To make a specific test of its robustness, I consider an interface charge of $\sigma=10^{13}{\rm cm}^{-2}$, and bulk that is either undoped or doped to $\rho_{\rm d}=10^{19} {\rm cm}^{-3}$, with other parameters as in GaN. \cite{GaN:parameters}

Fig.~\ref{fig:Fig3} displays the result of the robustness test, which  show that for this large charge, the general solution is a much better approximation for the potential when bulk is doped, that when it is undoped. 
This result can be understood as follows: As the first excited state gets occupied, it localizes and significantly increases the effective screening length. 
With doping, the excited states instead tend to delocalize, and their contribution to the negative charge density partly cancels with the background doping. 
The complete screening of the interface charge by the ground state is therefore equivalent with the approximation that the charge density of the excited states cancels with that of the bulk donors, which is impossible for zero or tiny donor density. 
The ground state energy does not typically agree well with that of the general solution.
The insert shows that for a fairly small interface charge $\sigma=10^{12} {\rm /cm^2}$ and low temperature, 
the general solution  agrees well with the solution of the full SP equation yields virtually identical potentials.
This agreement also verifies the accuracy and robustness of the numerical solver used and presented in Ref.~\onlinecite{berland:thermal}.

\begin{figure}[t]
\includegraphics[width=8cm]{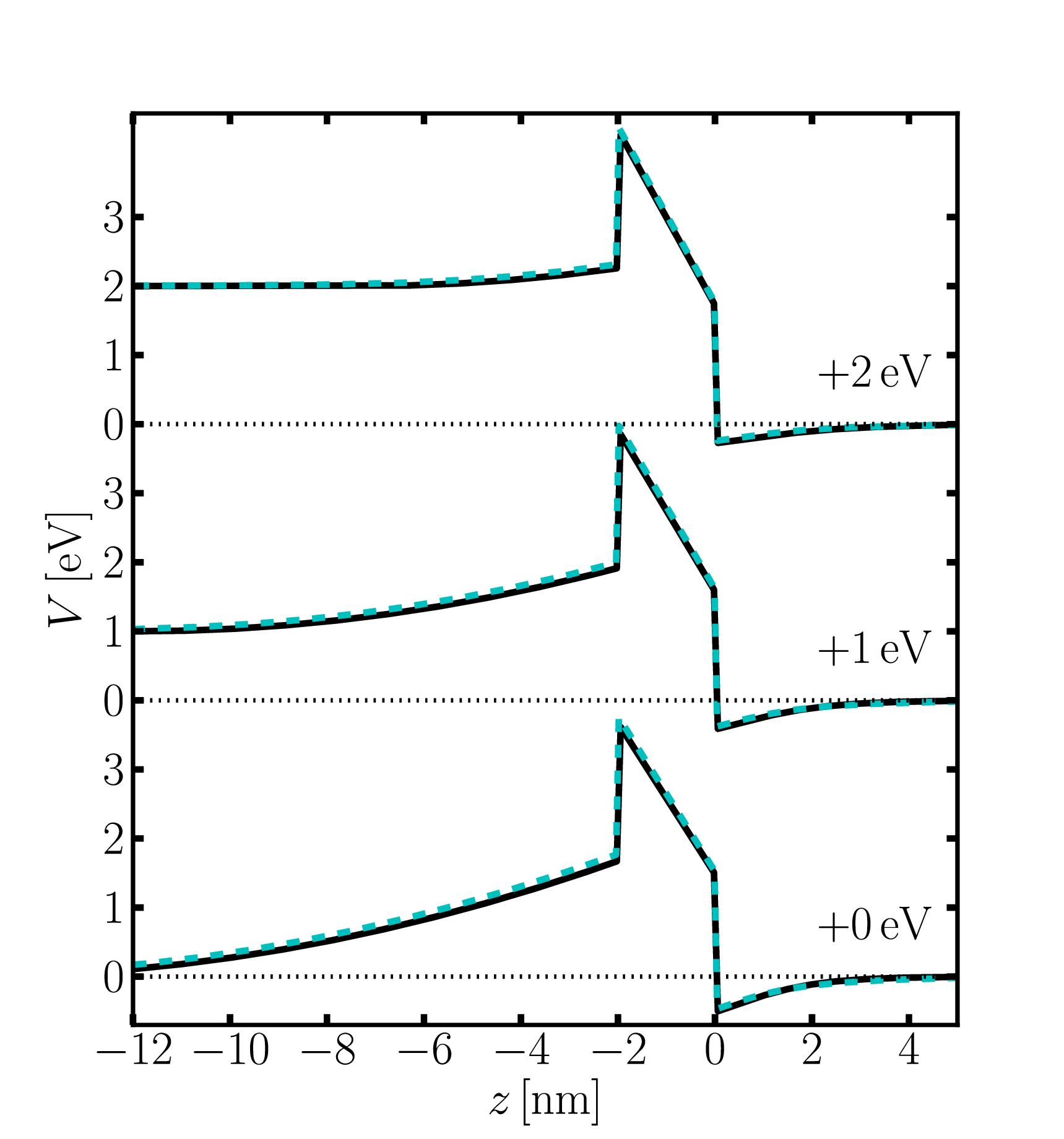}
\caption{Potential profile for AlN barrier sandwiched between doped ($\rho_d=10^{19} {\rm cm}^{-3}$) GaN cladding layers obtained from a SP calculation (dashed [cyan] curve)
with a step-function for the effective Fermi level and using an approximative analytical form (full [black] curve) for different biases. }
  \label{fig:Fig4}
 \end{figure}

Finally, Fig.~\ref{fig:Fig4} illustrates the usefulness of the approximative analytical form $U_{\rm app}$ given in Eq.~(\ref{eq:app}). It shows the potential profile for a system of a two nanometer wide AlN barrier sandwiched between GaN cladding layers at different bias $V_{\rm bias}$ with bulk doping $\rho_d$ as before. 
The full curves give the results of a simple model based on the general solution, while the dashed, the result of a full SP calculations using a step-function for the effective Fermi level. 
Such profiles are often displayed only for zero bias, perhaps because of the computational complication introduced by a nonconstant Fermi level.
To the left of the barrier, a depletion layer forms,\cite{berland:thermal,Hermann:AlN} which in the simple model is accounted for by a homogeneous charge density equalling the donor density $\rho_d$ with a depletion length of $L_{\rm dep}$.  
The analytical approximation for the potential profile $\gamma U_{\rm app}(x/\lambda)$ describes the the inversion layer at the right, with an energy $\gamma$ and length scale $\lambda$ that depends on the charge of the 2DEG $\sigma_{\rm 2DEG}$. The bias $V_{\rm bias}$ over the structure determines this charge:
\begin{equation}
\sigma_{\rm 2DEG}\frac{L_{\rm dep}/2+\lambda_{\rm 2DEG}(\sigma)<x>}{\eps_{\rm GaN}} +\left(\sigma_{\rm 2DEG}+\sigma_{\rm pol}\right) \frac{L_{\rm bar}}{\eps_{\rm AlN}}
\label{eq:mod}\,,
\end{equation}
where $\sigma_{\rm pol}$ is the charge stemming from the spontaneous and piezoelectric effects. Charge neutrality gives $\rho_{\rm d} L_{\rm dep}=\sigma_{\rm 2DEG}$. The full potential profile follows from simple electrostatics.
The result using this model agrees well with that of the SP calculation, which shows that the general, system-independent, solution can provide a quick and fairly accurate account of how the polarization in AlN/GaN heterostructures influence the potential profiles. 

In summary, the SP equation for a charged hard wall has, in the limit of complete screening by the ground-state, been expressed as a general, dimensionless equation. 
It leads to a simple relation between the charge at an interface and the conduction band bending. The approximative analytical expressions based on constraints stemming from the exact solution provide a convenient tool for obtaining the potential profile and charge density of heterostructures with large interface charges, such as for AlN/GaN structures. 
This could aid the design of intersubband devices in these materials.

%The simple model assumes that the depletion layer can be modelled as a homogenous charge $\rho(z)=\rho_d$ for $-\left(L_{\rm dep}+L_{\rm bar}\right)<z<-L_{\rm bar}$, while in the inversion layer $(z>0$, $\rho(z)=  \phi(z/\lambda) 
 
I thank P. Hyldgaard and T. G. Andersson for helpful discussions.
SNIC is acknowledged for supporting my participation in the National Graduate School in Scientific Computing (NGSSC). Financial support from Vinnova (banebrytende IKT).
\bibliographystyle{apsrev}
\bibliography{cw}
\end{document}